\documentclass[nofootinbib,aps,amsfonts,superscriptaddress]{revtex4}
\usepackage{graphicx,amsmath,amssymb,bm}
\usepackage{amsfonts,amssymb,amscd,amsmath}
\usepackage{graphicx}

\def\beq{\begin{equation}}
\def\eeq{\end{equation}}
\def\bea{\begin{eqnarray}}
\def\eea{\end{eqnarray}}
\def\eq#1{{Eq.~(\ref{#1})}}
\def\fig#1{{Fig.~\ref{#1}}}

\newcommand{\bas}{\bar{\alpha}_s}
\newcommand{\as}{\alpha_s}
\newcommand{\Lb}{\left(}
\newcommand{\Rb}{\right)}
\newcommand{\nn}{\nonumber}

\newcommand{\h}{\frac{1}{2}}
\newcommand{\un}{\underline}
\begin{document}


\title{ On the Nuclear Modification Factor at RHIC and LHC}
\author{Andrey Kormilitzin}
\affiliation{Department of Particle Physics, Tel Aviv University, Tel Aviv 69978, Israel}
\author{Eugene Levin}
\affiliation{Department of Particle Physics, Tel Aviv University, Tel Aviv 69978, Israel}
\affiliation{Departamento de F\'\i sica, Universidad T\'ecnica Federico Santa Mar\'\i a, Avda. Espa\~na 1680,
Casilla 110-V,  Valparaiso, Chile } 
\author{Amir H. Rezaeian}
\affiliation{Departamento de F\'\i sica, Universidad T\'ecnica Federico Santa Mar\'\i a, Avda. Espa\~na 1680,
Casilla 110-V, Valparaiso, Chile } 
\date{\today}
\begin{abstract}
We show that pQCD factorization incorporated with 
pre-haronization energy-loss effect naturally leads to flatness of
the nuclear modification factor $R_{AA}$ for produced hadrons at high transverse momentum
$p_T$. We consider two possible scenarios for the pre-hadronization:
In scenario 1, the produced gluon propagates through dense QCD medium and loses
energy. In scenario 2, all gluons first decay to quark-antiquark pairs
and then each pair loses energy as propagating through the medium. We show
that the estimates of the energy-loss in these two different models
lead to very close values and is able to explain the suppression of
high-$p_T$ hadrons in nucleus-nucleus collisions at RHIC. We show
that the onset of the flatness of $R_{AA}$ for the produced hadron in
central collisions at midrapidity is about $p_T
\approx 15$ and $25$ GeV at RHIC and the LHC energies, respectively.
We show that the smallness ($R_{AA}<0.5$ ) and the high-$p_T$ flatness
of $R_{AA}$ obtained from the $k_T$ factorization supplemented with the
Balitsky-Kovchegov (BK) equation is rather generic and it does not strongly depend
on the details of the BK solutions. We show that energy-loss effect
reduces the nuclear modification factor obtained from the $k_T$ factorization about $30\div 50\%$ at moderate $p_T$.

\end{abstract}


\maketitle
                       

\date{\today}

 

\section{Introduction}
\begin{figure}[t]
\includegraphics[width=14cm,height=8cm]{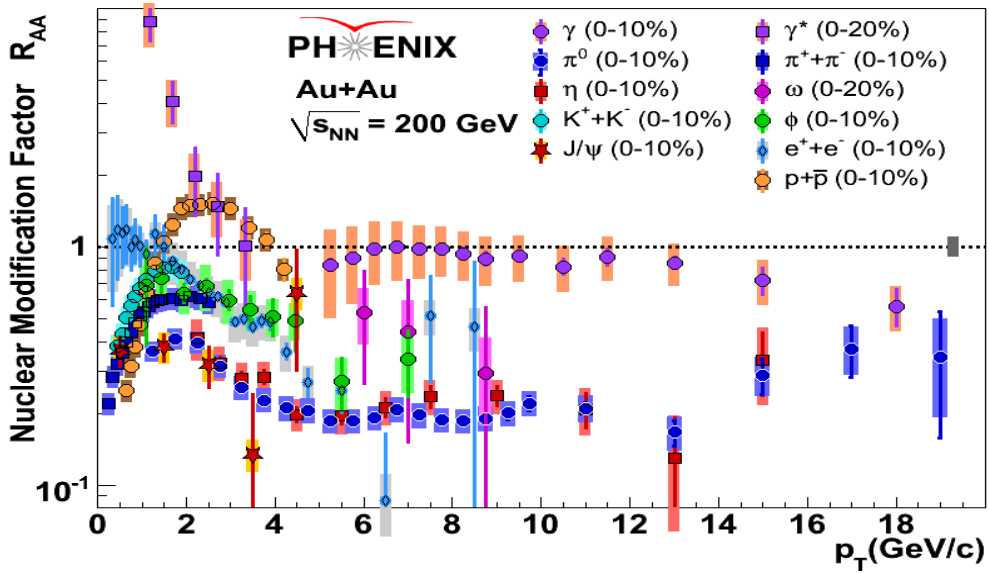}
\caption{The NMF for hadrons and direct photon versus transverse momentum $p_T$ in central Au+Au collisions at RHIC.
 The plot is taken from Ref.~\cite{NMF-n} from the PHENIX
 collaboration. }
\label{NMF}
\end{figure}

The nuclear modification factor (NMF) $R_{AA}$ is defined as
\beq \label{NMF1}
R_{AA}\,\,\equiv\,\,
\,\,\,\,\frac{1}{N_{coll}}\,
\,\frac{\frac{d^2 N_{AA }}{d y d^2 p_\perp}}{\frac{d^2 N_{NN}}{d y d^2 p_\perp}},
\eeq
 where $N_{coll}$ denotes the number of binary collisions in
 nucleus-nucleus (AA) collisions.  The measurements in $Au+Au$
 collisions at RHIC showed that the value of $R_{AA}$ is small and
 constant up to high transverse momentum of produced hadrons in
 central collisions \cite{NMF,BRAHMS0,BRAHMS1,BRAHMS2,PHENIX1}, see
 \fig{NMF}. At first sight such a behavior at high-$p_T$ contradicts
 the perturbative QCD (pQCD) factorization theorem
 \cite{FT1,FT2,FT3,FT4,col}. Accordingly to the factorization theorem, the
 inclusive gluonic-jet production cross-section with a large
 transverse momentum $p_T$ is proportional to $A^2
 \sigma_{\mbox{hard}} x\,G^2_p\Lb x ,p_T\Rb$ which leads to $R_{AA}\to 1$
 (see Sec. III for the details). On the other hand in the Color
 Glass Condensate (CGC) approach due to gluon saturation and the
 appearance of a new dimensional scale, saturation momentum $Q_s$
 \cite{GLR,hdQCD,MV,BK,JIMWLK}, there is a priori no reason to expect
 the pQCD factorization theorem to be valid. Indeed, it has been
 proven \cite{KTINC,KTFP1,KTFP2,KTFP3,KTFP4,KTFP5} that the 
 factorization theorem can be replaced by the $k_T$ factorization
 \cite{KTF1,KTF2,KTF3,KTF4} for scatterings of dilute-dilute or
 dilute-dense system of partons (like deep inelastic scattering with
 nuclei or scattering of two virtual -photons). For the case of
 scatterings of dense-dense systems when we have three scales: two
 saturation momenta for two nuclei and the transverse momentum of
 produced jet, the $k_T$ factorization has not been proven yet. However, for
 understanding of the NMF behavior, it is enough to discuss the
 behavior of the inclusive hadron production at transverse momentum
 $p_T$ larger than both saturation scales. In this kinematic region,
 the $k_T$ factorization works \cite{KTINC}. Having this in mind we
 can assume that the $k_T$ factorization works in all kinematic regions
 for scattering of dense-dense systems.

Based on the
 $k_T$ factorization, the inclusive production for gluon (jet) in $AA$ collisions at
 midrapidity can be calculated from the following equation, 
\beq \label{CGCIN}
\frac{d \sigma}{d y \,d^2 p_{T}}\|_{y = 0},\,= \,\frac{2C_F}{\as 2 (2\pi)^3}\,\frac{1}{x^2_\perp}\int d^2  b \,\,d^2  B \,\int^{+ \infty}_{- \infty}d  z \,e^{- z} \, J_0\Lb e^{\h z}\,x_\perp\Rb\,\, \nabla^2_z N_G\Big( z ; b \Big)\,\,
\, \nabla^2_z N_G\Lb z ; |\vec b-\vec B| \Rb,
\eeq
where  $p_T$ and $y$ are the transverse momentum and rapidity of the produced gluon, 
with notations $z \,=\,\ln\Lb r^2 Q^2_s\Rb$, $x_\perp= p_T/Q_s$ and $N_G = 2 N - N^2$.
We defined $C_F=(N^2_c-1)/2N_c$ where $N_c$ denotes the number of colors and $\alpha_s$ is the strong coupling.  
The forward dipole-nucleus amplitude $N$ can be obtained via solving 
the Balitsky-Kovchegov equation (BK)\cite{BK}. It is seen that
\eq{CGCIN} has $x_\perp$-scaling behavior in the kinematic region that the forward dipole amplitude $N$ has the geometric-scaling property.


Using \eq{CGCIN} we can re-write the NMF in the following form:
\beq \label{NMFT}
\mbox{NMF}\,\,\equiv\,\,\frac{1}{A^2}\,\frac{S^2_A}{S^2_p}\,\frac{{\mathcal{T} } \Lb x_\perp\Rb}
{{\mathcal{T}} \Lb x_\perp \frac{Q_{s,A}}{Q_{s,N}}\Rb}, 
\eeq
where function $\mathcal{T}$ is defined using \eq{CGCIN}. The area of
interaction (denoted by $S_A$ and $S_p$ for $AA$ and $pp$ collisions)
is well-defined for nucleus-nucleus scatterings while is rather
uncertain for proton-proton interactions \cite{LR1,LR2}. In the above
for simplicity we assumed that the area of interaction in both $AA$
and and $pp$ collisions can be factorized.

It turns out that the NMF calculated from Eq.~(\ref{NMFT}) is small $R_{AA}<0.5$ (see Sec IV). \eq{CGCIN} has the geometric
scaling behavior which is valid in the saturation region \cite{BALE}
and can be extended to a wider region outside the saturation domain
\cite{IIM}. The geometric scaling behavior is violated for $ z^2\,> 2
\ln \Lb Q^2_s(Y)/Q_s(0)\Rb $ where $Y\,=\,\ln(1/x)$ \cite{IIM}.
Notice that we have $2
\ln \Lb Q^2_s(Y)/Q_s(0)\Rb$ instead of $8\ln \Lb Q^2_s(Y)/Q_s(0)\Rb$
since $d^2 \sigma/d^2 p_t \propto N^2_G \propto N^4$ when $N$ is not
small. Therefore, for $z \gg
\sqrt{2 \ln \Lb Q^2_s(Y)/Q_s(0)\Rb}$ we expect that the scaling behavior will be broken and the inclusive cross section for jet production for both nucleus-nucleus and proton-proton collisions will be proportional to $\as(p^2_T)/p^4_T$ and consequently we have $R_{AA}\to 1$. 
Using the above equation and the saturation scale estimated in
Ref.~\cite{KLN} at RHIC energies, one may expect that only for $ p_T
\leq\,3 \div 4 \, Q_s$, the NMF to be different from unity. But a
slight glance at
\fig{NMF} shows that this is not the case for inclusive hadrons production for a wide range of $p_T$ while it
apparently holds for the production of direct photon \cite{pho,RS,pho1}. Therefore,  
the small value of the NMF at high $p_T$ may stem from large
distance interactions in medium and non-perturbative QCD should be
invoked to calculate such interactions \cite{MU1}. The data on
$R_{AA}$ for J/$\Psi$ production supports this conclusion \cite{PHENIXPSI,PHENIXPSI1,STARPSI,KOPPSI}.

In this paper, we show that the CGC predicts considerable suppression
of the NMF at low-$p_T$ both at RHIC and the LHC energies. We will
show that based on the CGC prescription at low-$p_T$, the NMF
slowly increases and then flattens at moderate $p_T$. Based on 
pQCD, we will then estimate, the onset of the flatness of the NMF at
large $p_T$.  The CGC approach is based on classical gluon fields and
evolution in leading log(1/x) approximation. In such
approximations the energy-loss in processes of gluon emissions is
neglected.  The systematic approach within pQCD framework
\cite{MU1,BDMS,ZA1,ZA2,WW,HH-f,Rev-d} shows that such emissions alone is not able to
describe the experimental values for the NMF and variety of
non-perturbative approaches have been developed
\cite{MU2,HKKKY,GUB,LRW,EL1,EL2,EL3,EL4,Rev-d}. In this paper, we introduce a simple 
approach for calculating the energy-loss effect due to
pre-hadronization which is able to describe RHIC data at high-$p_T$. At low-$p_T$,
energy-loss is less important and we show that the CGC prescription
\eq{NMFT} gives rather good description of RHIC data at small
Bjorken-$x$.

Our approach is based on two main ingredients: (i) the $k_T$
factorization \cite{KTINC,KTFP1,KTFP2,KTFP3,KTFP4,KTFP5} which takes
into account the rescatterings of the produced dipole (or gluon) in the final
state; and (ii) the energy loss due to the hadronization of the produced dipole (or gluon). 

In the next section, we calculate the energy-loss effect within two
different pre-hadronization pictures. In Sec. III, we discuss the
large $p_T$ behavior of the NMF for hadron production at RHIC and the
LHC.  Sec. IV is devoted to present our numerical results and
comparison with the experimental data. As a conclusion, in Sec. IV we
highlight our main results.

\section{Pre-hadronization models}
Hadronization, unfortunately, could be treated mostly
phenomenologically due to our lack of understanding of
non-perturbative QCD. There have been several approaches to describe
hadronization processes \cite{Rev-d,MU2,HKKKY,GUB,LRW,EL1,EL2,EL3,EL4}. 

Here, we propose a simple picture for hadronization hoping that the
phenomenological uncertainties can be reduced to few parameters which
can be then determined from experiment.  In our approach, a high
energy heavy ion collision can be simplified to three successive
stages: 1) parton production 2) pre-hadronization and 3) fragmentation
of the produced parton into hadron. At the first stage, we have
partons scattering and recombination effects which can be described
within the CGC approach. The CGC gives the description of the initial
wavefunction of projectiles and the multiple-parton production within
a time scale of the order of $1/Q_s$ after collisions.  In the second
stage, the produced dipoles propagate, emit gluons and interact with
the produced QCD medium and lose energy.  As we have already
mentioned, the emission of gluons and interactions with other dipoles
have been taken into account within the $k_T$ factorization if the
sizes of dipoles are small enough to trust the perturbative QCD
approach. In the CGC approach most of produced dipoles have sizes of
the order of $1/Q_s \,\ll\,1/\mu$ where $\mu$ is the soft scale.  The
energy loss for such dipoles can be taken into account by the BK
equation with a kernel that includes the energy conservation. The
estimates of Ref. \cite{KOLE} shows that these corrections cannot
explain the experimental data leading only to 20$\div$30\% suppression
of the NMF.  We believe that the main source of the energy loss at
this stage of process is originated from hadronization of the slow
dipoles that can be created during interaction of the fast dipole with
the slow remnants of the nuclear target or with the fast remnants of
the projectile nucleus. In the lab. frame one of the nucleus is at
rest and the produced dipole can interact with a nucleon in the target
or projectile elastically or inelastically. Elastic interaction does
not break the nucleon and does no lead to any energy loss. However, in
inelastic interactions nucleons are broken, and due to hadronization,
the remnants of the nucleon carry $(1 - z_h)$ fraction of energy
leading to $z_h$ energy loss of the fast dipole. The produced hadrons
do not interact since the density of the nucleons at fixed impact
parameters are small in the nuclei (maximum 2 nucleons for gold
). Finally, the sizes of produced dipoles during interaction with
nuclei reaches the soft scale ($1/\mu$) and due to hadronization the
energy of the fastest hadron becomes $z_h$ times smaller. Notice, that
we assume for simplicity that both hadronization stages have the same
$z_h$
 
In our description of hadronization, we  neglect
the fragmentation process and instead for simplicity and
clarity resort to the Local Parton-Hadron Duality (LPHD) principle \cite{LR1,LR2,KLN,ee},
namely we assume that the final-state hadronization is a soft process and cannot
change the direction of the emitted radiation\footnote{The same idea
was used in the KLN \cite{KLN} and the LR \cite{LR2} approaches which describes the rapidity
distribution of heavy-ion collisions data in a wide range of
energies. This scheme also describes experimental data from $e^+e^-$
annihilation into hadrons \cite{ee} and inclusive hadron productions in $pp$
collisions including the recent LHC data \cite{LR1}.}. Then, the key question is
to obtain the fraction of energy of the produced partonic system (in
the pre-hadronization stage) which is carried away by hadron, the
so-called $z_{A,h}$ parameter. We have recently shown
\cite{LR1} that the data for the inclusive hadron production in $pp$
collisions in a wide range of energies can be described if we assume
that the energy loss in the process of fragmentation of the produced
gluon into hadron reaches $z_h
\approx \h$.  In the following section we calculate the value of $z_A$ in the presence
of a dense QCD medium within two different pre-hadronization pictures and we will show
that both schemes give rather similar results.

\subsection{ Quark-antiquark pre-hadronization model}
\begin{figure}[t]
\includegraphics[width=10cm]{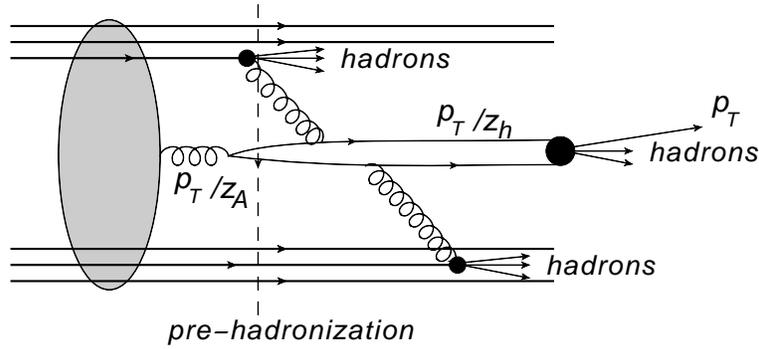}\\
\caption{ Inclusive production of hadrons in nucleus-nucleus collision through the stage of pre-hadronization.}
\label{qqeq}
\end{figure}

As a first model, in spirit of all available hadronization models,
we assume that hadronization passes a stage in which all gluons first go
to quark-antiquark pairs (pre-hadronization)(see \fig{qqeq}). Each
pair then decays into hadron in the same way for hadron-hadron,
hadron-nucleus and nucleus-nucleus interactions. In such a decay, 
the fastest hadron carries $z_h$ fraction of energy of the pair.

\begin{figure}[t]
\includegraphics[width=14cm]{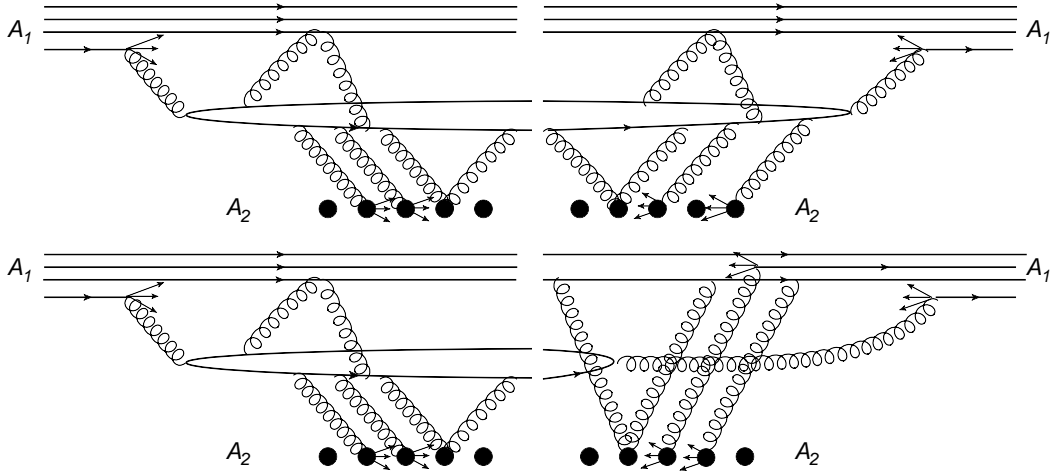}\\
\caption{ Inclusive production of quark-antiquark pair (dipole) in nucleus-nucleus collisions.
}
\label{qqprod}
\end{figure}

In nucleus-nucleus interactions, the process of quark-antiquark pair
creation is depicted in \fig{qqprod}. In heavy ion collisions, the
probability for a gluon \footnote{It is well known that at high energies the most produced partons are
gluons. However, at RHIC energies the contribution of quarks, perhaps,
are not negligible. However, in our approach similar to Ref.~\cite{KKT}, we
assume that gluons give the most important contribution at the highest
RHIC energy as well. The main reasons behind this assumption are: the main
contribution for the NMF factor measured at RHIC, stems from the
momentum of the jet $p_T = 8 - 20 \,\text{GeV}$ for $z = 1/2$. The typical
values of $x = 2 p_T/W/z$ turns out to be smaller than $0.1$ for
$W=200\,\text{GeV}$ while the quarks become essential at $x \geq 0.2$.
Moreover, our main goal here is not to describe the experimental data
but to answer the question whether it is possible to have the value of
the NMF much smaller that 1 within the framework of the QCD
factorization theorem. }
 with an average transverse momentum
$Q_{s1}$ to decay into quark-antiquark ($q\bar q$) pair at
impact-parameter $b$ with a relative transverse momentum denoted by
$\vec k\equiv \un k~(\un k^2\equiv k^2)$ has the following
form\footnote{The derivation is based on the approaches suggested in
Refs.~\cite{KTINC,KOV,KOP1,KOP2,RS}.} \cite{KLNT,TUCH,KTQQ},
\bea \label{MF1}
P^{q\bar q}_{AA}\Lb y, b, k\Rb  &=&\frac{\as\,Q_{s1}^2}{8 \pi^4}
\int d^2r\int d^2r'\, e^{-i\frac{1}{2} \un k\cdot(\un{r}-\un{r'})}\,\Big\{\frac{1}{2} \frac{\un{r}\cdot \un{ r'}}{rr'}\,K_1(r Q_{s1})\,
K_1(r' Q_{s1})+K_0(r Q_{s1})K_0(r'Q_{s1})\Big\}\, \nn\\
 &\times & \frac{8\,C_F}{\pi^2\as}\,
 \bigg\{ \frac{1}{r^2}\left(1-   e^{-\frac{1}{8}\un{r}^2Q_{s1}^2}\right) \left(1-   e^{-\frac{1}{8}\un{r}^2Q_{s2}^2}\right)+  
\frac{1}{r'^2} \left(1-   e^{-\frac{1}{8}\un{r'}^2Q_{s1}^2}\right)
\left(1-   e^{-\frac{1}{8}\un{r'}^2Q_{s2}^2}\right)\,\nn\\
& & ~~~~~~~~~~~~~~~~~~~~~~~~~~~~~~ -\,\,
\frac{1}{(\un{r}-\un{r'})^2}\left(1-   e^{-\frac{1}{8}(\un{r}-\un{r'})^2Q_{s1}^2}\right) \left(1-   e^{-\frac{1}{8}(\un{r}
-\un{r'})^2Q_{s2}^2}\right)    
   \bigg\}\, .
\eea
The factor in the first curly bracket is the distribution function
$\Psi( G \to q \bar{q} (r) \,\Psi^*( G \to q \bar{q} (r')$ \,that
describes the production of $ q \bar{q}$ dipoles by gluon with size
$r$ ($r'$) in the amplitude and complex conjugated ampplitude,
respectively, before the passage through the nuclei (see
\fig{qqprod}). This gluon has typical transverse momentum
$Q_{s1}$. The formula is written in the laboratory frame where the
nucleus $A_2$ is at rest. \eq{MF1} is derived as a sum of all possible
inelastic interactions. For example, for one inelastic interaction
with nucleus $A_2$ the contribution to \eq{MF1} has the following
form,
\beq\label{MF2}
\int d  z_1 e^{ - \h \sigma(Y,r)\,\rho z_1}\,\rho\,\sigma_{in}   e^{ - \h \sigma(Y,r)\,\rho ( 2R_{A_2} - z_1)}\,e^{  - \h \sigma(Y,r)\,\rho 2 R_{A_1}}\,\,e^{ - \h \sigma(Y,r')\,\rho ( 2R_{A_2} - z_1)}\,e^{  - \h \sigma(Y,r')\,\rho z_1},
\eeq
 where $\rho$ is the nucleon density and $R_{A_i}$ is the radius of the
nucleus $A_i$. \eq{MF2} indicates that a dipole with a transverse size $r$
interacts inelastically with the nucleon located at point $x_1$ in
nucleus $A_2$ and exponents provide that no inelastic interactions
occur with other nucleons in both nuclei in the amplitude. In the
complex conjugate amplitude the same happens with a dipole with a transverse
size $r'$. Note that for simplicity we assumed that the nuclear profile is cylindrical $T(b)\approx 2 R_A$ with $b^2\le R_A^2$.

The dipole-nucleon cross-section can be written as 
\begin{equation} \label{MF3}
\sigma \Lb Y, r \Rb\,=\,\frac{4\pi^2\as}{N_c}\,\int \,\frac{d^2 l}{2
  \pi l^2}\,\Lb 1 - e^{ i\,\un{r} \cdot\,\un{l}}\Rb\,\Lb 1 - e^{- i\,\un{r} \cdot\,\un{l}}\Rb\,\,\phi\Lb Y,  l^2\Rb\,\,\xrightarrow{\un{r} \cdot\,\un{l}\ll 1}\,\,
\frac{2\pi^2\as}{N_c}\,r^2\,\int d l^2 \phi\Lb Y,  l^2\Rb,  
\end{equation}
where $\phi$ is the unintegrated gluon density, $Y$ is the
rapidity of the dipole. In the right side of the above equation we
have kept only the dominant contribution relevant at large transverse momentum of the produced
dipole in the DGLAP approximation. Using Eq.~(\ref{MF3}), one can simplify Eq.~(\ref{MF2}) by the
following replacement \cite{KLNT,RS}:
\begin{equation} \label{SIG}
\sigma \Lb Y, r \Rb\,\rho\,T(b)\approx \sigma \Lb Y, r^2 \Rb\,\rho\,2 R_A\,\,\to   r^2 Q^2_{si}/8, 
\end{equation}
where $Q_{si}$ is the gluon saturation momentum in the nuclei $i$ and can also depend on the impact-parameter $b$. In the same fashion, for
inelastic dipole-nucleon cross-section we have the following expression
\bea \label{MF4}
\sigma_{in} \Lb Y, r,r' \Rb\,&=&\,2\,\frac{4\pi^2\as}{N_c}\,\int \,\frac{d^2 l}{2
  \pi l^2}\,\Lb 1 - e^{ i\,\un{r} \cdot\,\un{l}}\Rb \Lb 1 - e^{ -i\,\un{r'} \cdot\,\un{l}}\Rb \,\phi\Lb Y,  l^2\Rb\,\,\xrightarrow{\un{r} \cdot\,\un{l}\ll 1}\,\,
\frac{2\pi^2\as}{N_c}\, 2\,\un{r}\cdot\un{r'}\,\int d l^2 \phi\Lb Y,  l^2\Rb \nn, \\
\sigma_{in} \Lb Y, r, r'\Rb\,\rho\,2 R_A\,\,&\to & 2 \un{r}\cdot\un{r'}  Q^2_{si}/8.  
\eea
The inelastic cross-section can be generally written in a form of
convolution of two amplitudes $\sigma_{in} =
\sum_n A\left( r; \{k_i\}\right) \times A^*\left( r';\{k_i\}\right) $ with sum over
produced $n$-partons with momenta $k_i$. \eq{MF4} describes the inelastic
cross-section for the process in which the dipole of size $r$ in the
amplitude interacts with nucleon while in the
complex-conjugated amplitude, the dipole with size $r'$ interacts with the same nucleon as in the amplitude.

Note that \eq{MF1} integrated over the impact-parameter $b$ gives the
inclusive cross-section for $q\bar q$-pair with a relative
transverse momentum $k$. It incorporates the first stage of the process,
namely the parton production (shown as a dark blob in \fig{qqeq}).
Here, we assumed that the produced quark-antiquark pair is moving in
the classical gluon field, neglecting the evolution in the rapidity
interval $y$. We believe that this assumption will not be essential
in estimating the energy loss.

In calculating the average energy loss, we assume that in each
inelastic interaction, the dipole energy decreases $z_h$ times due to
the pre-hadronization process.  For elastic interaction we also
introduce factor $z_{h,el}$ which describes the fraction of energy
that is carried by the fastest hadron. For simplicity we assume that
$z_{h}=z_{h,el}$.  For elastic scattering the energy loss emerges only
on the last stage, namely quark-antiquark (or gluon) hadron
transition.  An additional source of the energy loss in nucleus
collisions stems from the inelastic interaction of $q\bar q$-pairs (or
gluons). In this process, soft gluons (slower than propagating $q\bar
q$-pair or gluon) are produced. Then, these gluons produce hadrons due
to hadronization and these hadrons carry $(1 - z_h)$ energy of
colliding gluon.  Due to this process, the propagating gluon loses
$z_h$ fraction of energy per each inelastic interaction. 
Therefore, in order to calculate the energy loss in a nucleus we need to find
the probability to have the inelastic collisions with $i$ nucleons in
a nucleus ($P_i$) and multiply it by the energy loss in each such
collisions. In other words, 
\beq \label{AVZNEW}
<z_A> \,\,=\,\,z_h P_0 + \sum_{i=1}\,z^i_h P_i, 
\eeq
where we define the probability $P_i = \sigma_{i, in}/\sigma_{in}$
with $\sigma_{i,in}$ being the contribution of inelastic \,$i$ \,
collisions to the total inelastic cross section $\sigma_{in}$. The
latter is given by Eqs.~(\ref{MF1},\ref{MF2}). Therefore, in order to calculate the
resulting average energy-loss of the dipole propagating through the
medium, in Eqs.~(\ref{MF1},\ref{MF2}), one needs to find $\sigma_{i,in}$ and
multiply every inelastic collision  by the factor $z_h$. 
From Eqs.~(\ref{MF3},\ref{MF4}) one can see that each factor
$\un{r}\cdot\un{r'} $ is associated with the inelastic
interaction.
Therefore, this multiplication can be carried out by the
following replacement in \eq{MF1},
\beq \label{MF5}
(\un{r}-\un{r'})^2\,=\,r^2 + r'^2 - 2 \un{r}\cdot \un{r'}\,\,\to\,\,r^2 + r'^2 - 2\,z_h\, \un{r}\cdot \un{r'}.
\eeq
The fact that factor $\un{r}\cdot\un{r'} $ is associated with the
inelastic cross-section in Eq.~(\ref{MF4}) for $P^{q\bar q}_{AA}\Lb y, b, k\Rb$
can be understood on general grounds. For example, the inelastic
dipole cross-section at two-gluon exchange level is $\sigma_{in} \propto r
r' $. Notice that all other terms in expansion of $P^{q\bar q}_{AA}\Lb
y, b, k\Rb$ are of the order of $r^2$ and $r'^2$.

The average energy loss can be
then obtained from the following expression
\beq \label{MF6}
\langle z_h \rangle_A\,\,\equiv\,\,z_A\,\,=\,\,\Big( \Lb z_h\,-\,1\Rb\, 
P^{q\bar q}_{AA}\Lb y, b, k;z_h=0\Rb \,\,+P^{q\bar q}_{AA}\Lb y, b, k; z_h\Rb\Big)\,\, \Big{/}P^{q\bar q}_{AA}\Lb y, b, k;  z_h = 1 \Rb.
\eeq
In the above equation, the term $P^{q\bar q}_{AA}\left( y, b,
k;z_h=0\right)$ describes the process of diffraction dissociation in
which dipoles scatter only elastically and the term $P^{q\bar
q}_{AA}\left( y, b, k; z_h\right) - P^{q\bar q}_{AA}\left( y, b, k;z_h=0\right)$
corresponds to the contribution of all inelastic processes with
additional factor $z_h$ for each rescattering.

Therefore, the first term in \eq{MF6} only takes into account the
processes of hadronization of the quark-antiquark with rapidity $y$
that passes through the medium without an inelastic interaction.  The
second term in
\eq{MF6} is responsible for hadrons production in the entire kinematic region of rapidity $ y $. Note
that \eq{MF6} is based on the assumption that if no energy-loss occurs
in a single interaction $z_h=1$ then no energy-loss will follow in
multiple-interactions namely $z_A=1$, and if the parton loses all its
energy in a single interaction $z_h=0$ then naturally in
multiple-interactions we have also $z_A=0$. The NMF \eq{NMFT}
incorporating the pre-hadronization energy-loss effect, can be then
rewritten in the following form,
\beq \label{NMFTF}
\mbox{NMF}\,\,=\,\,\frac{1}{A^2}\,\frac{S^2_A}{S^2_p}\,\frac{{\mathcal{T}} \Big(p_T\Big{/}\Lb z_A\,Q_{s,A}\Rb\Big)}
{{\mathcal{T}} \Big( p_T \Big{/}\Lb z_h\,Q_{s,N}\Rb\Big)}.
\eeq
In \eq{NMFTF} we use the fact that the transverse momentum of the emitted gluon in $pp$ or $AA$ collisions is equal to \cite{LR1}, 
\beq \label{PTZ}
p^2_T\Lb \mbox{gluon} \Rb \,\,\,=\,\,p^2_T\Lb \mbox{hadron} \Rb/z^2_{h,A} \,\,+\,\,k^2_T\Lb\mbox{intrinsic}\Rb,
\eeq
where $ k_T\Lb\mbox{intrinsic}\Rb$ is the average intrinsic transverse
momentum of the gluonic jet and since its value is small, it can be 
neglected. 

\subsection{Gluon pre-hadronization model}

\begin{figure}[t]
\includegraphics[width=10cm]{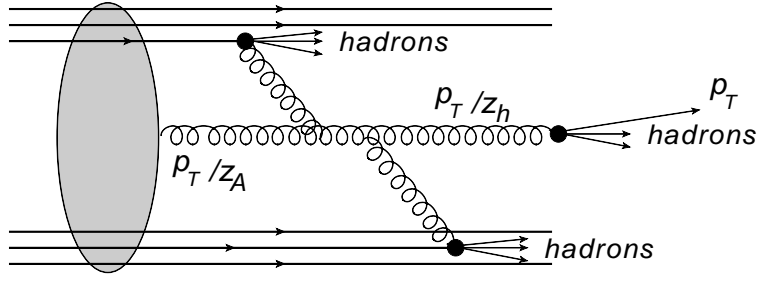}\\
\caption{ Inclusive production of hadrons in nucleus-nucleus collision due to gluon decay.}
\label{nmfeq}
\end{figure}

In the second model, we consider a case that the produced gluon decays
into hadrons after propagating through the nucleus (see \fig{nmfeq}).

\begin{figure}[t]
\includegraphics[width=14cm]{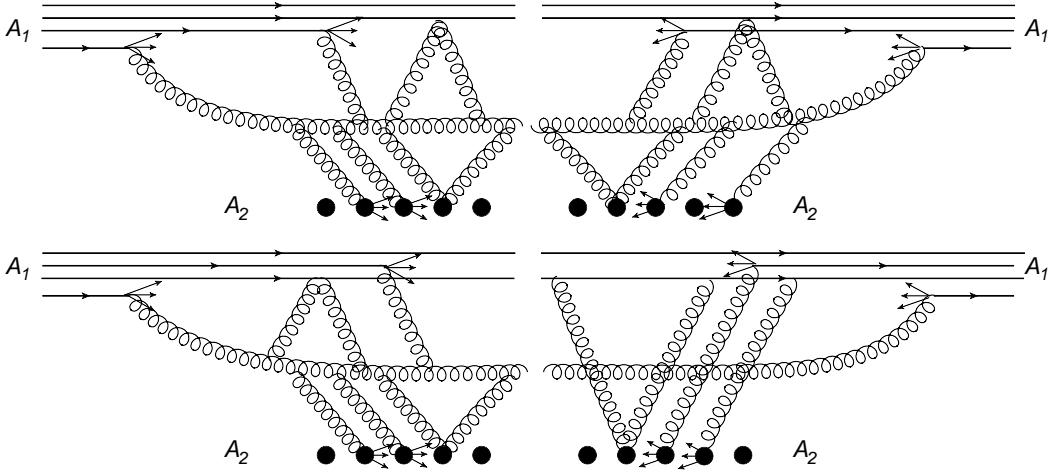}\\
\caption{ The passage of the produced gluon in nucleus-nucleus collision.}
\label{glprod}
\end{figure}

Using the approaches suggested in Refs.~\cite{KTINC,KOV,KOP1} and
summing all gluon exchange diagrams shown in \fig{glprod} and 
neglecting again the small-$x$ evolution in the rapidity interval between
the produced gluon and the target, the probability of gluon production in $AA$ collisions 
with a transverse momentum $k$ can be written in the following
form,

\bea \label{MF51}
P^G_{AA}\Lb y, b, k\Rb  &=&\frac{\as\,Q_{s1}^2}{8 \pi^4}
\int d^2r\int d^2r'\, e^{-i\frac{1}{2} \un k\cdot(\un{r}-\un{r'})}\,\bigg\{ K_2\Lb k r\Rb\, K_2\Lb k r'\Rb\,J_0\Lb k r\Rb\,J_0\Lb k r'\Rb\,\bigg\}\, \nn\\
 &\times & \frac{8\,C_F}{\pi^2\as}\,
 \bigg\{ \frac{1}{r^2}\left(1-   e^{-\frac{1}{4}\un{r}^2Q_{s1}^2}\right) \left(1-   e^{-\frac{1}{4}\un{r}^2Q_{s2}^2}\right)+  
\frac{1}{r'^2} \left(1-   e^{-\frac{1}{4}\un{r'}^2Q_{s1}^2}\right)
\left(1-   e^{-\frac{1}{4}\un{r'}^2Q_{s2}^2}\right)\,\nn\\
& & ~~~~~~~~~~~~~~~~~~~~~~~~~~~~~~ -\,\,
\frac{1}{(\un{r}-\un{r'})^2}\left(1-   e^{-\frac{1}{4}(\un{r}-\un{r'})^2Q_{s1}^2}\right) \left(1-   e^{-\frac{1}{4}(\un{r}
-\un{r'})^2Q_{s2}^2}\right)    
   \bigg\}\, .
\eea
In the derivation of the above equation, we assume that the gluon with
transverse momentum $k$ is emitted by the dipole with transverse size smaller
than $1/k$. Indeed, all rescattering of the gluons emitted by dipoles with larger sizes are already included in
\eq{CGCIN} \cite{KTINC}.  In this case the emission of gluon from interaction with a nucleon can be considered as
a decay of the nucleon to two colorless dipoles. This can be better
conceived at large $N_c$ limit where an adjoint (gluon) dipole
can be decomposed into two fundamental (quark) dipoles.  These two
dipoles penetrate through the medium with the same cross section. The
wave function of such a gluon has been calculated in
Ref.~\cite{GW}. Therefore, this approach differs from the
quark-antiquark pre-hadronization model mainly by the value of the
dipole cross section: $\sigma_G\Lb Y,r^2\Rb \approx\,\,2
\sigma_{dipole}$.  In other words, the main differences between
\eq{MF51} and \eq{MF1} is that in \eq{MF51} the value of the
saturation momentum is two times larger than in \eq{MF1}.

Notice than in general, we are dealing here with rescatterings of two
dipoles with different sizes. For example, in the dipole-nucleus
scattering amplitude at the first stage of the process, the incoming
dipole with size $x_{12}=| \vec{r}_1 - \vec{r}_2|$ emits a gluon with
a coordinate $\vec{r}$ and this process at $N_c
\gg 1$ can be considered as a decay of the dipole $x_{12} $ to two
dipoles $x_{13} = |\vec{r}_1 - \vec{r}|$ and $x_{23} = |\vec{r}_2 -
\vec{r}|$. These two dipoles take part in interactions with a
nucleus. However, in the inclusive cross-section we integrate over the
momenta that carry quark (antiquark) with coordinates $r_1$ and $r_2$.
Therefore, this cross-section does not depend on these coordinates and
both can be put equal to zero. That is the reason why the inclusive
cross-section here can be viewed as a propagation of dipole with size $r$
in the amplitude and with size $r'$ in the complex conjugated
one.


In order to calculate the average energy loss of the produced gluon at
the pre-hadronization stage $\langle z_h \rangle_A$, we again follow the steps in
Eqs.~(\ref{MF5},\ref{MF6}) by replacing $P^{q\bar q}_{AA} \to P^G_{AA}$. The
corresponding NMF can be then obtained by \eq{NMFTF}.

It is worthwhile mentioning that in \eq{NMFTF} we take into account both the first stage, namely $q
\bar{q}$-pair (or gluon) production (the dark blobs in Figs.~\ref{qqeq}, \ref{nmfeq}), and also the energy-loss effect due to the passage of the produced $q \bar{q}$-pair (or gluon)
through the medium (see Figs.~\ref{qqprod}, \ref{glprod}). In a sense, we unfolded the $k_T$
factorization and made the corresponding corrections due to the
pre-hadronization energy-loss effect since the hadronization leads to
its violation.

\begin{figure}[t]
\includegraphics[width=6cm]{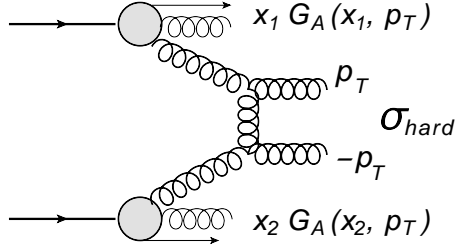}\\
\caption{ The gluon production accordingly to the factorization theorem at large $p_T$.}
\label{nmfft}
\end{figure}


\section{Large $p_T$ behavior of NMF}

\begin{figure}[t]
\begin{tabular}{c c}
\includegraphics[width=7cm]{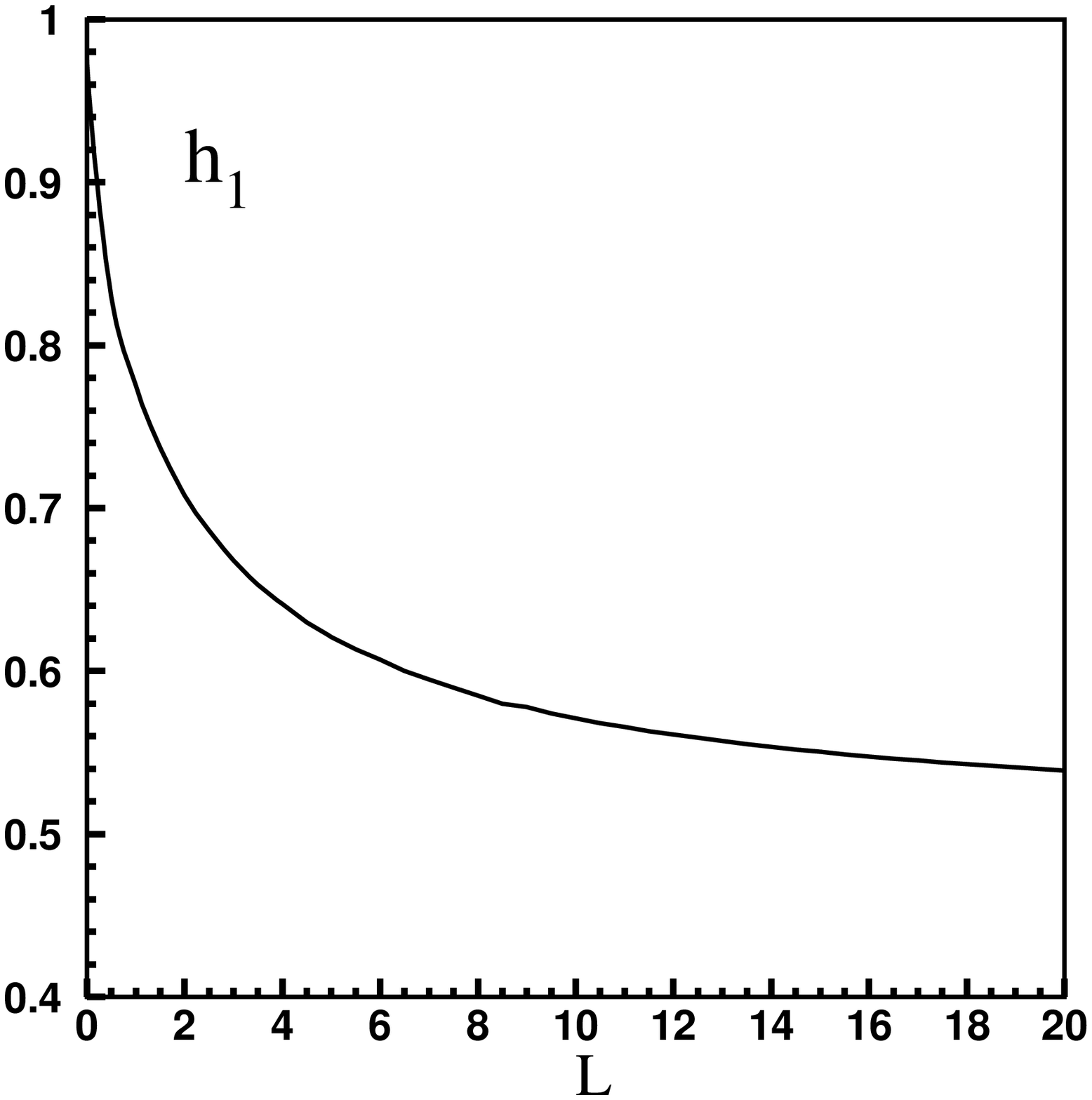}& 
\includegraphics[width=7cm]{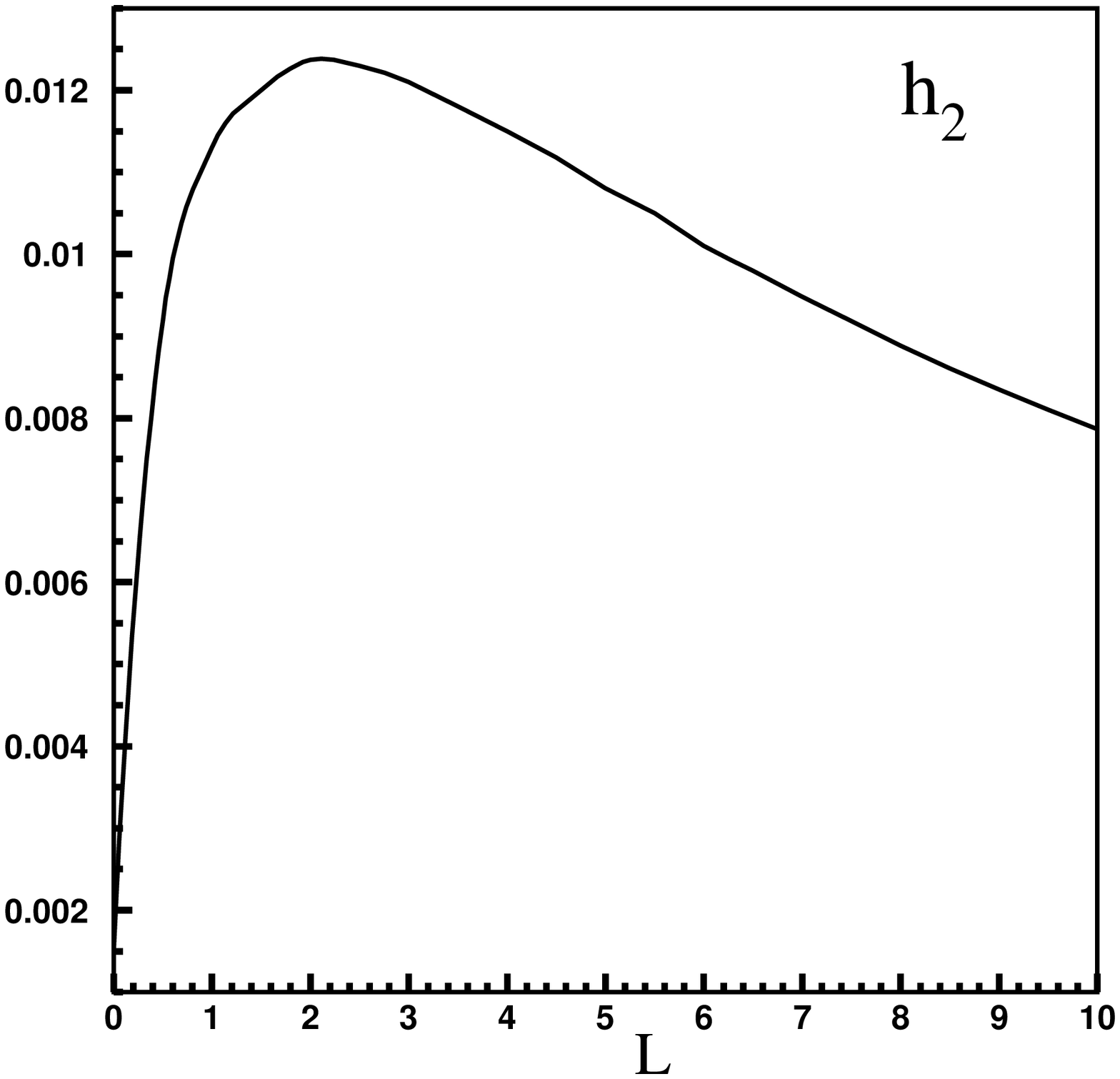}
\end{tabular}
\caption{ Function $h_1$ and $h_2$ defined in Eqs.~(\ref{LPT241},\ref{LPT242}) versus $L$. }
\label{hf}
\end{figure}

At very large $p_T$, the pQCD factorization theorem
\cite{FT1,FT2,FT3,FT4,col} can be used. The production of gluon-jet with
transverse momentum $p_T$ (see \fig{nmfft}) in nucleus-nucleus
collisions can be simply written as
\bea \label{LPT1}
\frac{d \sigma_{AA}}{d y, d^2 p_T}|_{y=0}\,\,&=&\,\,A^2 \frac{d \sigma_{pp}}{d y, d^2 p_T}\,\,= \,\, \sigma_{\mbox{hard}}\,
x_1 G_A\Lb x_1= 2 p_T/\sqrt{s}, p_T\Rb\,x_2 G_A\Lb x_2= 2 p_T/\sqrt{s}, - p_T\Rb,\\
&=&\,\,
A^2  \frac{\as^2(p_T)}{p^4_T } \, x_1 G_p\Lb x_1= 2 p_T/\sqrt{s}, p_T\Rb\, x_2 G_p\Lb x_2= 2 p_T/\sqrt{s}, - p_T\Rb
\,\to\,A^2  \frac{\as^2(p_T)}{p^4_T } \Lb p^2_T/Q^2_0 \Rb^{2 \gamma}\nn, 
\eea
where $xG_P$ (or $xG_A$) is the gluon structure function for the
proton (or nucleus) and $Q_0$ is a separation scale so as for $p_T >
Q_0$ one can use perturbative QCD. In the above expression,
$\sigma_{\mbox{hard}}$ is the perturbative partonic cross section
computable up to a given order in $\alpha_s$, and $\gamma$
denotes the anomalous dimension. It is well-known that in the
leading-order for $x \to 1$ we have $\gamma \to \as$. Therefore, at
very large $p_T\, \gg\, Q_s $ but $ p_T \,\ll\, \sqrt{s}$, the NMF
obtained via Eqs.~(\ref{NMF1},\ref{PTZ},\ref{LPT1}) is equal to
\beq \label{LPT2}
\mbox{NMF}\,\,\xrightarrow{\sqrt{s} \,\gg p_T\, \gg\, Q_s } \,\,
\frac{\as^2\Lb p_T/z_A\Rb\left( p_T/z_A\right)^{4\as(p_T/z_A)}}{\as^2\Lb p_T/z_h\Rb\left( p_T/z_h\right)^{4\as(p_T/z_h)}}\times \Big(\frac{z_A}{z_h}\Big)^4.  
\eeq 
The value of the NMF can be then immediately estimated at large $p_T$
via \eq{LPT2} by only knowing the value of $z_A$, $z_h$ and also
$\Lambda_{QCD}$ which appears in the running strong-coupling. For the
running coupling $ \alpha_s$, we employ the scheme used in Ref.~\cite{LR1} at the leading-order. We
postpone the numerical discussion to the next section, and in the rest
of this section, we address another important question, namely
at which $p_T$ the expression given in \eq{LPT2} can be used. Let us for sake of
simplicity work in double log approximation with the anomalous
dimension $\gamma =
\bas/\omega$ where $\omega$ is Mellin image of $\ln(1/x)$ and  $\bas=\alpha_s N_c/\pi$ . In such an approximation, the nuclear gluon structure
function up to a numerical constant $\mathcal{K}$ can be written as
\cite{LT}
\bea \label{LPT20}
\frac{x G_A\Lb x= 2 p_T/\sqrt{s}, p_T\Rb}{p^2_T} \,\,&=&\,\,\mathcal{K} \,\exp\Lb \sqrt{ 4 \bas \ln(1/x) \,\ln\Lb p^2_T/Q^2_0\Rb}\,\,-\,\,
\ln\Lb p^2_T/Q^2_0\Rb\,\,+\,\,l_A\Rb\,\,\nn, \\
&=&\,\,\mathcal{K} \,\exp\Lb \sqrt{ L \,\ln\Lb p^2_T/Q^2_0\Rb}\,\,-\,\,
\ln\Lb p^2_T/Q^2_0\Rb\,\,+\,\,l_A\Rb,
\eea
where $L \,=\, 4 \bas \ln(1/x)\,\,=\,\,\ln\Lb Q^2_s\Lb A; x\Rb/Q^2_s\Lb A; x=x_0\Rb\Rb$ and $l_A = (1/3)\ln A$.
The equation for the saturation scale has the form \cite{LT},  
\beq \label{LPT21}
L\, l_s\,\,=\,\,( l_s - l_A)^2,
\eeq
with a solution, 
\beq
l_s\Lb L,l _A\Rb\,\equiv\, \ln\Lb Q^2_s\Lb A; x\Rb/Q^2_0\Rb\,\,=\,\,\frac{L}{2}\,+\,l_A\,\,
\sqrt{\Lb \frac{L}{2}\Rb^2 \,+\,L\,l_A}.
\eeq
One can see that $G_A$ given by \eq{LPT20} at $Q^2=Q^2_0$ is proportional to $A^{1/3}$.
Introducing a new variable $\mathcal{Z} = \ln\Lb p^2_T/Q^2_s\Lb A; x\Rb\Rb$ one can rewrite \eq{LPT20} in the following form
\beq \label{LPT22} 
\frac{x G_A\Lb x= 2 p_T/\sqrt{s}, p_T\Rb}{p^2_T} \,\,=\,\,\mathcal{K} \,\exp\Lb H\Lb \mathcal{Z},L,l_A\Rb\Rb, 
\eeq
with a notation, 
\beq
 H\Lb \mathcal{Z},L,l_A\Rb\,\,=\,\,\sqrt{L\,\Lb \mathcal{Z} + l_s\Lb L,l_A\Rb\Rb}\,\,-\,\,\mathcal{Z}\,\,-\,\,l_s\Lb L,l_A\Rb\,\,+\,\,l_A.
\eeq
Expanding $H\Lb z,L,l_A\Rb$ at small $z$, we have
\beq \label{LPT23}
H\Lb \mathcal{Z},L,l_A\Rb\,\,=\,\,-\,h_1\Lb L,l_A\Rb\,\mathcal{Z}\,\,+\,\,h_2\Lb L,l_A\Rb\,\mathcal{Z}^2\,\,\,+\,\,\,{\cal O}\Lb \mathcal{Z}^3\Rb, 
\eeq
with
\bea 
h_1\Lb L,l_A\Rb\,\,&=& \,\,\,\,1\,\,-\,\,\frac{L}{2 \sqrt{ \frac{L^2}{2} \,+\,L\,l_A \,+\,L\,\sqrt{l^2_A \,+\,\Lb \frac{L}{2} + l_A\Rb^2}}},\label{LPT241}\\
h_2\Lb L,l_A\Rb\,\,&=& \,\,\,\,\frac{L^2}{2 \sqrt{2} \Lb L \Lb L + 2 l_A + \sqrt{L^2 + 4 L l_A + 8 l_A^2}\Rb \Rb^{3/2}}.  \label{LPT242}
\eea
The function $h_1$ and $h_2$ versus $L$ for the gold are shown in \fig{hf}.
It is seen that at large values of $L$, we have $h_1 \to 1/2$ and $h_2
\to 1/8L$ in accordance with finding in Ref.~\cite{IIM}. Using the KLN saturation model parametrization, 
the value of $L$ is about  $L \approx 1 $ and $1.7$ for RHIC and the LHC energies, respectively 
\cite{KLN}. The condition \cite{IIM}
\beq \label{LPT3}
h_2\,\mathcal{Z}^2 \,>\,1,
\eeq
characterizes that $\mathcal{Z}$ is large and the scattering amplitude is far
away from the saturation domain, out of the region where we have the
geometric scaling behavior at large values of $L$.  Indeed,
the first term in \eq{LPT23} leads to the geometric scaling behavior
at large values of $L$ while the second term violates this behavior
even at large $L$. Therefore, the condition
given in \eq{LPT3} ensures that the gluon structure function defined
in \eq{LPT22} gives the perturbative form used in \eq{LPT1}.

\eq{LPT3} leads to $p_T >  40$ GeV for the RHIC energies and  to $p_T >  60$  GeV for the LHC energies for the produced gluon.
In terms of the transverse momenta of the produced hadron we expect that
for $p_T \geq 40\,z_A\approx 15$ GeV at the RHIC and for $p_T \geq
60\,z_A\approx 25$ GeV at the LHC, the NMF defined in 
\eq{LPT2} to be reliable.  Notice that if the pre-hadronization stage in $AA$ and $pp$ collisions
was the same, namely $z_A=z_h$, then the NMF defined in \eq{LPT2} was
identically equal to one $R_{AA}= 1$.  Therefore, the main source of
suppression of the NMF for hadrons at high-$p_T$ is due to further
energy loss of the produced gluons in $AA$ collisions compared to $pp$
collisions at the pre-hadronization stage, namely $z_A<z_p$. Notice
that on average $z_A$ and $z_p$ vary slowly with $p_T$
for the kinematic region of our interest here \cite{LR2}. 
Note also that the $p_T$-dependence in \eq{LPT2} mainly enters through the running of the strong-coupling. 
This justify the flatness of the NMF $R_{AA}$ over large
range of $p_T$. In the next section we show that the numerical value
obtained for $z_A$ from the pre-hadronization mechanisms introduced in
section II can indeed describe the observed suppression of hadrons at
large-$p_T$ at RHIC. For smaller values of $p_T$, \eq{CGCIN} may be
used with the dipole amplitude $N$ being in the region where we still
have the geometric scaling behavior for $N \equiv N(z)$ \cite{IIM}.


\section{Numerical results and comparison with experimental data}

In order to estimate the value of $\langle z_h \rangle_A$, we use the
KLN saturation parametrization \cite{KLN} with $Q_s=1.41$ GeV at
$\sqrt{s} = 200$ GeV for gold and we take $k = 2$ GeV as a typical
scale for soft interaction. The chosen soft interaction scale corresponds to the
value of the soft Pomeron slope $k \approx 1/\sqrt{\alpha'_{I\!\!P}}$
with $\alpha'_{I\!\!P}= 0.25 \,\text{GeV}^{-2}$ \cite{KOP1,KOP2,KOP3}. It turns out that both pre-hadronization models give very close values for $\langle z_h
\rangle_A/z_h$ : $0.766$  and $0.744$ for the quark-antiquark and the gluon pre-hadronization model respectively, at the RHIC energy 
$\sqrt{s} = 200$ GeV. The value of $z_h\approx 1/2$  was
obtained from a fit to the experimental data in $pp$ collisions for the average
transverse momentum of charged hadrons for a wide range of energies
\cite{LR1}.

\begin{figure}[t]
\includegraphics[width=10cm]{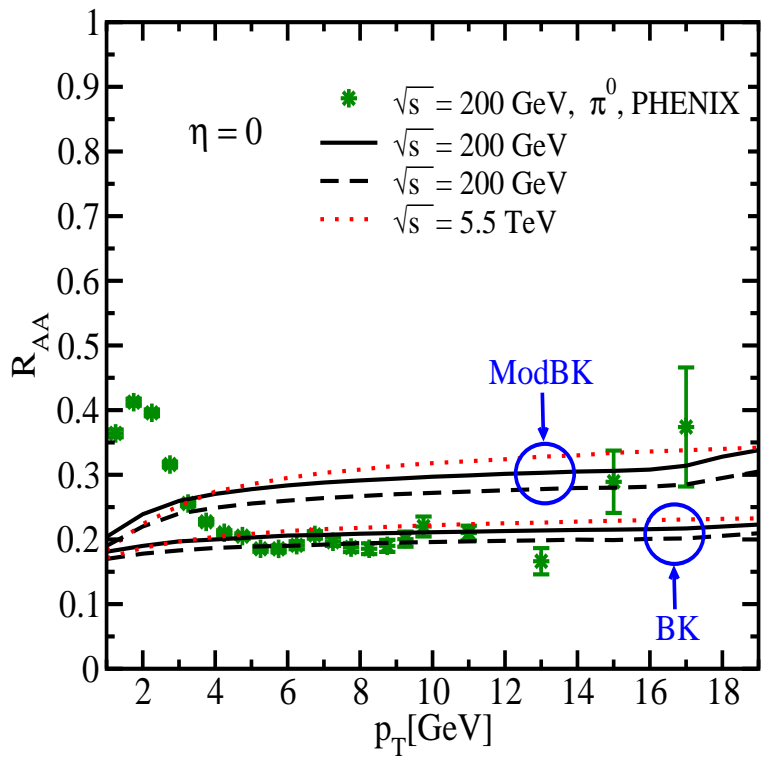}
\caption{The NMF for hadrons obtained by \eq{NMFTF} for $0-10\%$ centrality $Au+Au$ collisions at various energies. 
Function $\mathcal{T}\Lb x_\perp\Rb$ is calculated using the solution
of the BK and the modified BK equations.
The solid and dashed lines correspond to the quark-antiquark and gluon
pre-hadronization models, respectively. The experimental data is taken from Ref.~\cite{NMF}.}
\label{nmfest1}
\end{figure}

Again using the KLN parametrization for the nuclear saturation scale, we estimated the value of $\langle z_h
\rangle_A/z_h$  at the LHC energies, assuming $z_h \approx 1/2 $. At higher energies or larger
value of the saturation scale $Q_s$, one can approximate the integrand
of Eqs.~(\ref{MF1},\ref{MF51}) to its maximum value at $r = r \prime
\approx 1/Q_s$. This leads to a similar value for $z_A$ in both
pre-hadronization models at higher energies, since the over-all
factors in Eqs.~(\ref{MF1},\ref{MF51}) will be canceled out via
Eq.~(\ref{MF6}). Indeed, full numerical solution is in accordance with this observation and
we found that both pre-hadronization models result to a very similar value
for $z_A$ at the LHC energies, namely $\langle z_h \rangle_A/z_h \approx 0.82,~0.795$ at
$\sqrt{s} = 5.5,~2.75$ TeV, respectively.

Our results for the NMF for $0-10\%$ centrality in $Au+Au$ collisions using
the $k_t$ factorization \eq{NMFTF} supplemented with solutions of the BK
and the modified BK equation that preserves the energy conservation
\cite{KOLE} is shown in \fig{nmfest1}. In the modified BK equation (denoted by ModBK in \fig{nmfest1}), the next-to-leading order corrections to the BFKL
kernel are taken into account which leads to conservation of energy in
the framework of non-linear equation \cite{LKOLE}. We recall that here
we resort to the LPHD principle for the final state hadronization. But
the energy-loss effect in the pre-hadronization both in $pp$ and $AA$
collisions are effectively incorporated in the results shown in
\fig{nmfest1}. We believe that the difference between two
pre-hadronization models (introduced in Sec. II) in $AA$ collisions
gives reasonable estimates of the errors which are unavoidable in the
situation that we do not know the theory of pre-hadronization in a
dense medium. As we already pointed out, the differences between these
two pre-hadronization models at LHC is negligible. The energy-loss
effect reduces the NMF defined in \eq{NMFTF} about $30\div 50\%$.  Notice
that this reduction does not
depend on a given BK solution and incorporates the missing energy-loss effect into the $k_T$ factorization.

We should stress
that the numerical estimates in
\fig{nmfest1} was obtained from the BK equation with a simplified
kernel \cite{LT,KOLE}. Such estimates is less reliable for describing
the experimental data, even though, interesting enough, such a rough
approximation apparently is in agreement with RHIC data at high-$p_T$
within the errors.  Notice that in fact the validity of the $k_T$
factorization \eq{NMFTF} beyond the extended geometric-scaling
region at high-$p_T$ is questionable, see also Sec. I. An important
message here is that the smallness and the high-$p_T$ flatness of the
NMF obtained from \eq{NMFTF} and the BK equation is rather generic and
it does not strongly depend on the details of the BK solutions in our
interested kinematic region here, see also recent interesting paper by
Albacete and Marquet \cite{javier}. It is also seen from \fig{nmfest1}
that as one may expect, the onset of the flatness of the NMF at the
LHC energy occurs at higher $p_T$. \fig{nmfest1} indicates that the
energy-conservation constrain on the BK equation which is important at
higher $p_T$ tends to enhance the NMF. This is due to the fact that
the ModBK solution correctly incorporates the behavior of the
anomalous dimension at large Bjorken-$x$ by matching to the DGLAP value.

\begin{figure}[t]
\includegraphics[width=10cm]{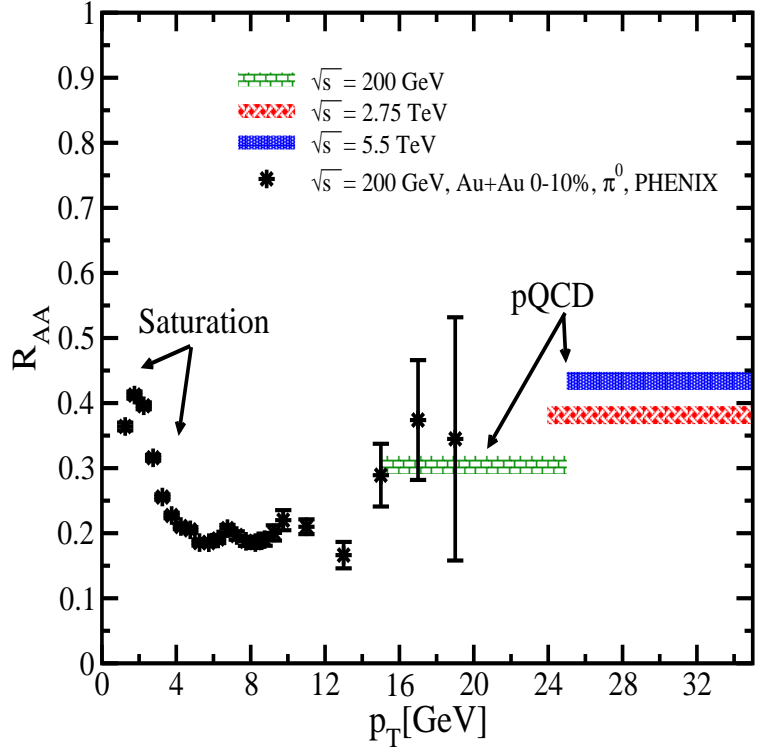}
\caption{ The NMF for hadrons obtained by \eq{NMFTF} at low $p_T$ and \eq{LPT2} at high $p_T$ for $0-10\%$ centrality $AA$ collisions at various energies. Function $\mathcal{T}\Lb x_\perp\Rb$ is calculated within the saturation (CGC)
approach \cite{LR1,LR2}. We also show pQCD results obtained in Sec. III. The experimental data is taken from
Ref.~\cite{NMF}.}
\label{f-r}
\end{figure}

In order to make more reliable prediction for the NMF for the neutral pion $\pi^0$, we next
re-calculate the NMF from \eq{NMFTF} by using the impact-parameter
dependent CGC saturation model in $pp$ \cite{LR1} and $AA$ collision
\cite{LR2}.  It has been already shown that such a scheme describes HERA data at small-$x$ \cite{wat} and the hadron multiplicities in nucleus-nucleus collisions at RHIC \cite{LR2} and  provided correct predictions \cite{LR1} for the inclusive hadron production at the LHC. 
The importance of the inclusion of the impact-parameter dependence in
the saturation models and the BK equation has been addressed in
Refs.~\cite{LR1,LR2,SSS}.  Our results at RHIC and the LHC energies in $0-10\%$
centrality $Au+Au$ (at RHIC) and $Pb+Pb$ (at the LHC) collisions, is shown in Fig.~\ref{f-r}. We recall that \eq{NMFTF} has
$x_\perp$-scaling property which is valid only at low transverse
momentum roughly for $p\leq 3\div 4 Q_S$. In the saturation model
\cite{wat} used in Fig.~\ref{f-r}, the saturation scale varies very
slowly with energy and for $p_T = 1$ GeV at midrapidity for central
collisions we have $Q_s=0.76$ GeV for proton at $\sqrt{s}=5.5$
TeV. Therefore, the upper limit of $x_\perp$-scaling behavior in this
model is about $p_T\approx 3 $ GeV for $pp$ collisions (the reference
for the NMF) at $\sqrt{s}=5.5$ TeV at midrapidity. We therefore, only
show the CGC results coming from \eq{NMFTF} for $p_T<4$ GeV. 
Here, we only concentrate on the CGC results at rather 
low-$p_T$ where we can also give a good description of spectra in $pp$ collisions \cite{LR1}.
At the RHIC energy $\sqrt{s}=200$ GeV at mid-rapidity, for $p_T>2$ GeV, we
have $x>0.01$ which is beyond the validity of the CGC
prescription. Note also that the dipole model used here was obtained
from a fit to HERA data for virtuality $Q^2\in [0.25,
45]~\text{GeV}^2$. Therefore, this model is valid for $p_T\in [0.7,
6.7]$ GeV. Again, we employ the LPHD principle to connect
pre-hadronization to final hadronization stage. This leads to two
different over-all factors for the spectra in $pp$ and $AA$ collisions
which are fixed with the experimental data at lower energies
\cite{LR1,LR2}. The appearance of two different over-all factors for
the normalization is partly due to the fact that pre-hadronization
leads to different effective masses for the mini-jet in $AA$ and $pp$
collisions \cite{LR1,LR2}. It is seen from Fig.~\ref{f-r} that our
results without any free-parameter to adjust, is in accordance with
the PHENIX data at low-$p_T$, in contrast to the impact-parameter
independent BK solution shown in Fig.~\ref{f-r} with a zero mini-jet
mass.  At low-$p_T$, the energy-loss in medium is less important and
we have taken $z_A=z_h=1/2$ for $p_T<4$ GeV. Notice that the actual
value of $R_{AA}$ at low $p_T$ calculated via the $k_T$ factorization
may be subjected to about $15\div 25\%$ errors due to the uncertainties in
the hadron multiplicity obtained within the same framework in $AA$
collisions. Moreover, here the contribution of quarks was neglected
which is perhaps less justifiable at midrapidity at low-energy.

At high-$p_T$, we rely on the improved pQCD factorization result given
in
\eq{LPT2} (also shown in Fig.~\ref{f-r}). The only external parameter in the NMF defined in \eq{LPT2} is the parameter $z_A$ which incorporates the
energy-loss effect at the pre-hadronization stage. The value of the
parameter $z_A$ was already calculated in Sec. II (see also above) at
RHIC and the LHC energies.  As we already argued \eq{LPT2} is valid at
high transverse momentum of the produced hadrons for $p_T
\geq 40\,z_A\approx 15$ GeV at RHIC and for $p_T \geq 60\,z_A\approx
25$ GeV at the LHC energies. Our predictions for the NMF suppression
and the onset of the flatness of the NMF at high-$p_T$ both agree with
the PHENIX data.  As we already pointed out \eq{LPT2} naturally leads
to the flatness of the NMF at high $p_T$ since $p_T$ dependence mainly
enters via the running strong-coupling and varies very slowly with
$p_T$. In Fig.~\ref{f-r}, we also show our predictions both at low and
high $p_T$ at the LHC energies $\sqrt{s}=2.75$ and $5.5$ TeV. The
uncertainties in our formulation is less than $10\%$ which come mainly
from the normalization in
\eq{NMFTF}.
 The band in Fig.~\ref{f-r} indicates about $2\%$ theoretical error
 which also includes the discrepancies between two pre-hadronization
 schemes.  The predictions of other approaches at the LHC can be found in Ref.~\cite{other}.


\section{Conclusions}

In this paper first we developed a simple picture for calculating the
energy-loss effect.  In our approach, the energy-loss effect stems
from the process of the pre-hadronization. We showed that the
estimates of the energy-loss in two different models of the
pre-hadronization lead to very close values and is able to explain the
measured $R_{AA}$ at high transverse momentum of produced
hadrons at RHIC. We showed that the small value of the nuclear modification
factor at high-$p_T$ is mainly due to the energy-loss effect in the
pre-hadronization stage.

We also investigated the NMF obtained from the $k_T$ factorization
supplemented with solutions of the BK and the modified BK equation that
preserves the energy conservation. We showed that the smallness
($R_{AA}<0.5$ ) and the high-$p_T$ flatness of the NMF obtained from
\eq{NMFTF} and the BK equation is rather generic and it does not
strongly depend on the details of the BK solutions. We showed that the
modified BK solution which includes the energy conservation tends to
slightly enhance the NMF at high-$p_T$.  This is due to the fact that
the modified BK equation properly includes the anomalous dimension at
$x \approx 1$ which coincides with the anomalous dimension of the
DGLAP equation while the BK equation only reproduces its double log
limit. This indicates that one should be cautious as extrapolating
the $k_T$ factorization results to higher $p_T$ is less reliable if the
running anomalous dimension obtained from the BK solution does not
match to the DGLAP value at high-$p_T$.

We showed that at high-$p_T$, the pQCD factorization incorporated with
the pre-haronization energy-loss effect \eq{LPT2} results naturally to
flatness of the NMF. We obtained the onset of the applicability of
\eq{LPT2} to be for $p_T > 40$ GeV at RHIC and for $p_T > 60$ GeV at
the LHC for the produced gluon. Notice that in this kinematic region
the NMF for the produced jet should approach to one $R_{AA}^g \to
1$. However, the NMF of the produced hadron is very small
$R_{AA}^{\pi^0} < 0.5$ due to the pre-hadronization effect, namely
$z_A/z_p<1$. In terms of the transverse momenta of the produced hadron
we expect that this to be for $p_T \geq 15$ GeV at RHIC and for $p_T
\geq 25$ GeV at the LHC. 
As one can see from Figs.~\ref{nmfest1}, \ref{f-r} the hadronization
model is responsible for $30 \div 50 \%$ of the NMF value while $50
\div 70 \%$ of this value stems from the BK equation and $k_T$
factorization.

One of important check of our hadronization model will be the
estimates of the NMF in deuteron-nucleus collisions which will be
published elsewhere.  Here we would like only to mention that in our
approach, estimates for the NMF in proton-gold collision at the RHIC
energy $\sqrt{s}=200$ GeV at midrapidity gives only upto $10\%$
contribution due to the hadronization leading to the value of the NMF about $0.95 \pm 0.05$.

Finally, it is worth mentioning that the main assumption that in the process
of hadronization the fastest particle carries the final part of the
initial energy is a typical feature of all hadronic reactions as well
as the production of hadrons in DIS.

\begin{acknowledgments}
This work was supported in part by Conicyt Programa Bicentenario PSD-91-2006 and the Fondecyt (Chile) grants 1090312 and 1100648.

\end{acknowledgments}

\end{document}